# ALPHA DECAY FAVOURED ISOTOPES OF SOME SUPERHEAVY NUCLEI: SPONTANEOUS FISSION *VERSUS* ALPHA DECAY

O. V. KIREN, S. B. GUDENNAVAR[*] and S. G. BUBBLY

Department of Physics, Christ University, Bangalore-560 029, Karnataka, India
E-mail: shivappa.b.gudennavar@christuniversity.in

Spontaneous fission and alpha decay are the main decay modes for superheavy nuclei. The superheavy nuclei which have small alpha decay half-life compared to spontaneous fission half-life will survive fission and can be detected in the laboratory through alpha decay. We have studied the alpha decay half-life and spontaneous half-life of some superheavy elements in the atomic range Z = 100-130. Spontaneous fission half-lives of superheavy nuclei have been calculated using the phenomenological formula and the alpha decay half-lives using Viola-Seaborg-Sobiczewski formula (Sobiczewski *et al.* 1989), semi empirical relation of Brown (1992) and formula based on generalized liquid drop model proposed by Dasgupta-Schubert and Reyes (2007). The results are reported here.

*Key words*: Superheavy nuclei, Spontaneous fission, Alpha decay, Half -life.

## 1. INTRODUCTION

The synthesis of superheavy nuclei has received considerable attention in recent years with the advent of modern accelerators and suitable detectors [1-7]. The existence of long-lived superheavy nuclei is controlled mainly by the spontaneous fission and alpha decay processes. Bohr and Wheeler [8] described the mechanism of nuclear fission and established a limit $Z^2/A \approx 48$ for spontaneous fission, on the basis of the liquid drop model, beyond which nuclei are unstable against spontaneous fission. The alpha decay of superheavy nuclei is possible if the shell effect supplies the extra binding energy and increases the barrier height of fission [9-13]. Compared to alpha decay, the situation in spontaneous fission is very complex as there are large uncertainties existing in fission process, such as mass, charge number of two fragments, the number of emitted neutrons, released energy etc. Beta decay is another possible decay mode for the superheavy nuclei, but, since the beta decay proceeds *via* a weak interaction, the process is slow and less favoured compared to spontaneous fission and alpha decay. Beta stable nuclei having relatively longer half-life for spontaneous fission might decay *via* alpha emission. Therefore superheavy nuclei which have short alpha decay half-life compared to spontaneous fission half-life will survive fission and can be detected in the laboratory through alpha decay. In order to produce artificial superheavy nuclei, one needs the theoretical understanding of dominant mode of decay of superheavy nuclei. There have been some efforts made recently by many researchers in this direction [11, 14-21]. In this regard, we make an attempt to identify the long lived superheavy elements for spontaneous fission by comparing the calculated alpha decay half-life with the spontaneous fission half-life. The alpha decay half-lives of elements with Z=100 to 130 have been calculated using Viola-Seaborg Semi empirical relation [22], semi empirical formula of Brown [23] and generalized liquid drop model [24] while spontaneous fission half-lives have been calculated using the phenomenological formula proposed by Ren and Xu [25]. This may provide a very helpful insight to conduct experiments to realize the presence of superheavy nuclei.

---

[*] Corresponding author: S. B. Gudennavar

## 2. THEORY

The main modes of decay for superheavy nuclei are alpha decay and spontaneous fission. Alpha decay and spontaneous fission theoretically share the same phenomenon - the quantum tunnelling effect. There have been many theoretical studies on alpha decay and spontaneous fission half-life of heavy nuclei by many researchers [22-25]. We have used the Viola-Seaborg semi empirical formula [22], semi empirical formula proposed by Brown [23] and generalized liquid drop model proposed recently by Dasgupta-Schubert and Reyes [24] for studying the alpha decay half-lives.

The Viola–Seaborg semi-empirical relationship (VSS) with constants determined by Sobiczewski *et al*. [22] for alpha decay half-lives is given by

$$\log_{10}(T_{1/2}/\sec) = (aZ+b)Q^{-1/2} + cZ + d + h_{\log} \quad (2.1)$$

where the *Q*-value is in MeV and '*Z*' is the atomic number of a parent nucleus, *a*, *b*, *c* and *d* are constants and their values are, $a = 1.66175$, $b = -8.5166$, $c = -0.20228$, $d = -33.9069$. The quantity $h_{\log}$, the hindrance factor, in equation (2.1) accounts for the hindrances associated with odd proton and odd neutron numbers, whose values are, $h_{\log} = 0$ for even-even nucleus, $h_{\log} = 1.066$ for even-odd nucleus, $h_{\log} = 0.772$ for odd-even nucleus and $h_{\log} = 1.114$ for odd-odd nucleus. Once the *Q* values are calculated, the alpha decay half-lives can be estimated using the equation (2.1).

The semi empirical formula proposed by Brown [23] for determining the half-life of superheavy nuclei is given below,

$$\log_{10}(T_{1/2}/\sec) = 9.54(Z-2)^{0.6}/\sqrt{Q_\alpha} - 51.37 \quad (2.2)$$

where Z is the atomic number of parent nucleus, $Q_\alpha$ is in MeV.

Another formula for determining the half-lives of superheavy nuclei, proposed by Dasgupta-Schubert and Reyes [24] based on generalized liquid drop model, was obtained by fitting the experimental half-lives for 373 alpha emitters. The formula is given by

$$\log_{10}(T_{1/2}/\sec) = a + bA^{1/6}Z^{1/2} + cZ/Q_\alpha^{1/2} \qquad (2.3)$$

The coefficients are $a = -25.31$, $b = -1.1629$, $c = 1.5864$ for even-even, $a = -26.65$, $b = -1.0859$, $c = 1.5848$ for even-odd, $a = -25.68$, $b = -1.1423$, $c = 1.5920$ for odd-even and $a = -29.48$, $b = -1.113$, $c = 1.6971$ for odd-odd parent nuclei.

The spontaneous fission half-lives are calculated using the phenomenological formula proposed by Ren and Xu [25] and is given by

$$\log_{10} T_{1/2} = 21.08 + C_1 \frac{(Z-90-\nu)}{A} + C_2 \frac{(Z-90-\nu)^2}{A} + C_3 \frac{(Z-90-\nu)^3}{A} \\ + C_4 \frac{(Z-90-\nu)(N-Z-52)^2}{A} \qquad (2.4)$$

where $C_1 = -548.825021$, $C_2 = -5.359139$, $C_3 = 0.767379$ and $C_4 = -4.282220$, the seniority term $\nu$ was introduced taking the blocking effect of unpaired nucleon on the transfer of many nucleon-pairs during the fission process and $\nu = 0$ for spontaneous fission of even-even nuclei, $\nu = 2$ for odd A and odd-odd nuclei.

### 3. RESULTS AND DISCUSSION

A comparative study of alpha decay *versus* spontaneous fission is made for even-even, even-odd, odd-even and odd-odd superheavy nuclei in the atomic range Z = 100 to 130 using the semi empirical relations given in the last section. The masses and binding energies are available in the literature only upto Z = 130. The Q values are computed using experimental binding energies of Audi *et al.* [26, 27] and some masses are taken from table of KUTY [28].

Figures 1 depicts the comparison of the calculated alpha decay and spontaneous fission half-lives against mass number of even-even isotopes (for example, Z = 100, 102, 104 and 106) covering the atomic number range Z = 100 to 130. The isotopes $108^{264-270}$, $110^{268-276}$, $112^{272-280}$, $114^{274-284}$, $116^{278-290}$, $118^{280-294}$, $120^{284-300}$, $122^{288-304}$, $124^{292-312}$, $126^{293-316}$, $128^{296-321}$ and $130^{298-326}$ situated within the inverted parabola have the alpha decay channel as the dominant mode of decay, and are suitable to be synthesized and studied as these isotopes can be identified in the laboratory *via* alpha decay. Some of these isotopes (for *e.g.* $108^{264-270}$, $110^{268-276}$, $112^{272-280}$, $114^{274-284}$, $116^{278-290}$, and $118^{280-294}$ etc.) have been already synthesized experimentally and identified *via* alpha decay.

The alpha decay half lives calculated using all the three approaches (VSS, GLDM and Brown formulae) for the superheavy nuclei upto $Z = 116$ are compared with experimental half-lives in Table 1, which shows that they are in close agreement except in a few cases. From the table it is clear that the values from GLDM approach agree fairly well with the experimental values.

In Table 2, we compare the alpha decay (using GLDM approach) and spontaneous fission half lives of all isotopes starting from $Z = 100$ to $Z = 130$, which are found to live longer for spontaneous fission. From the table it is clear that alpha decay half-life is very much less than the spontaneous fission half-life for these isotopes and these can be identified in the laboratory *via* alpha decay.

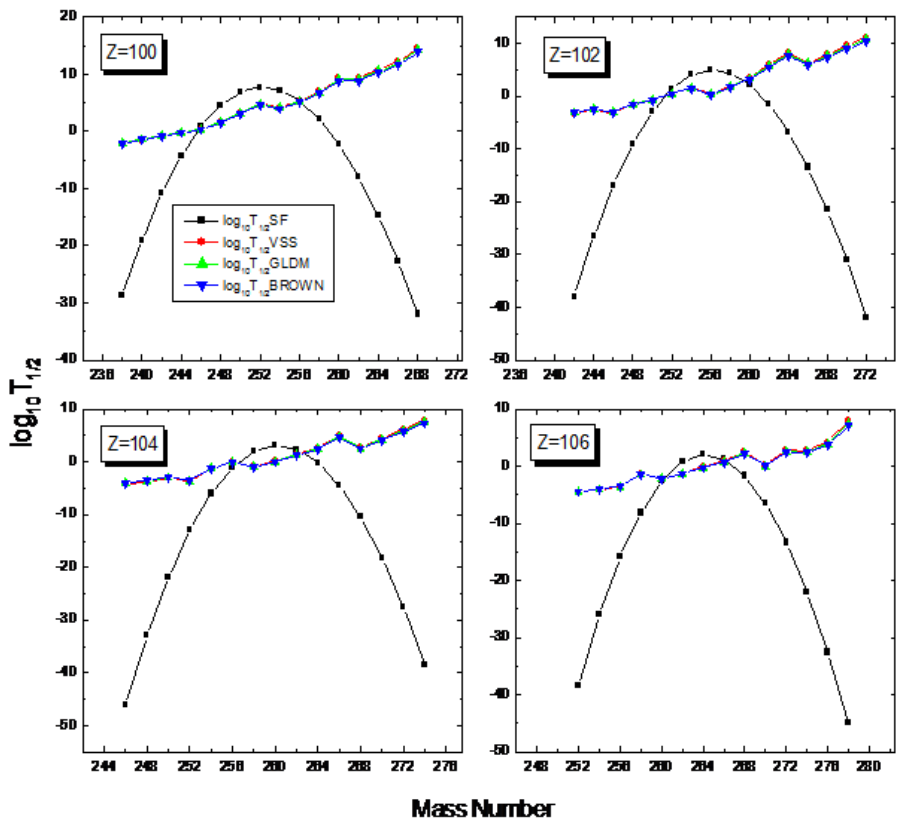

Fig. 1 – Comparison of alpha decay and spontaneous fission half lives for even-even isotopes with $Z = 100$, 102, 104 and 106.

In the case of fission process, the fissioning nucleus with sufficiently high excitation energy behaves like a structureless charged liquid drop. Therefore the probability of fission is governed by a parameter called fissility parameter that relates the Coulomb and surface energy of the nucleus. Fissility parameter is defined as the ratio of Coulomb energy of a spherical sharp surface drop to twice the nuclear surface energy and is given as,

$$X = \frac{Z^2/A}{50.883\left(1 - 1.7826\left(\frac{N-Z}{A}\right)^2\right)} \quad (2.5)$$

where $Z^2/A$ is the fissionability parameter, Z is atomic number, A is the mass number, and N is the neutron number of the nucleus.

Figure 2 represents the plot connecting the ratio of spontaneous fission to alpha decay half lives against fissility parameter for even-even, even-odd, odd-even, odd-odd isotopes. From the plot we see that for a given 'Z' value, the ratio of spontaneous fission to alpha decay half-lives increases with increasing fissility parameter, reaches a maximum and then decreases. The competition between spontaneous fission and alpha decay against the fissility parameter shows the same trend for even-even, even-odd, odd-even and odd-odd nuclides.

*Table 1*

Comparison of alpha decay half-lives calculated theoretically with the experimental values

| A | Z | $\log_{10}T_{1/2}$ | | | |
|---|---|---|---|---|---|
| | | Experimental | VSS | GLDM | Brown (1992) |
| 250 | 100 | 3.25 | 3.22 | 3.21 | 3.09 |
| 254 | 100 | 4.14 | 4.19 | 4.11 | 4.00 |
| 260 | 106 | −2.07 | −2.13 | −2.17 | −2.12 |
| 266 | 106 | 0.39 | 0.89 | 0.75 | 0.67 |
| 264 | 108 | −3.60 | −3.22 | −3.27 | −3.15 |
| 290 | 116 | −1.82 | −2.55 | −2.78 | −2.60 |
| 292 | 116 | −1.74 | −1.07 | −1.34 | −1.28 |
| 270 | 110 | −3.79 | −4.07 | −4.16 | −3.94 |
| 280 | 110 | 1.04 | 0.98 | 0.71 | 0.65 |
| 282 | 112 | −3.30 | −0.30 | −0.54 | −0.54 |
| 286 | 114 | −0.79 | −1.66 | −1.89 | −1.79 |
| 287 | 114 | −0.29 | 2.42 | 1.81 | 0.90 |
| 289 | 114 | 0.43 | 3.38 | 2.73 | 1.77 |
| 290 | 116 | −1.82 | −2.55 | −2.78 | −2.60 |
| 292 | 116 | −1.74 | −1.07 | −1.34 | −1.28 |
| 294 | 118 | −2.74 | −1.48 | −1.70 | −1.66 |
| 285 | 112 | 1.53 | 4.19 | 3.52 | 2.54 |
| 283 | 112 | 0.60 | 1.84 | 1.21 | 0.42 |
| 279 | 110 | −0.74 | 0.28 | −0.37 | −0.95 |
| 271 | 106 | 2.16 | 2.76 | 2.08 | 1.42 |
| 274 | 111 | −1.82 | −2.31 | −2.68 | −3.36 |
| 270 | 109 | −0.24 | −1.41 | −1.76 | −2.52 |
| 284 | 113 | −0.32 | -0.13 | −0.42 | −1.41 |
| 283 | 113 | −1.00 | −4.55 | −4.84 | −5.08 |
| 279 | 111 | −0.77 | −1.61 | −1.93 | −2.42 |

Figure 3 represents the plots connecting the spontaneous fission half-lives versus fissionability parameter ($Z^2/A$) for even-even, even-odd, odd-even, odd-odd isotopes. For a given atomic number Z, the spontaneous fission half-life increases with increasing fissionability parameter, reaches a maximum and then decreases with increasing fissionability parameter.

*Table 2*

Comparison of alpha decay (using GLDM) and spontaneous fission half lives for isotopes with Z = 100 to 130.

| Z | A | (Alpha decay) $\log_{10}T_{1/2}$ (sec.) | (Spontaneous fission) $\log_{10}T_{1/2}$ (sec.) |
|---|---|---|---|
| 100 | 252 | 4.81 | 7.72 |
|  | 253 | 4.87 | 11.29 |
| 101 | 254 | 3.26 | 9.63 |
|  | 255 | 3.14 | 9.55 |
| 102 | 256 | 0.41 | 5.02 |
|  | 257 | 2.32 | 7.96 |
| 103 | 258 | 1.37 | 6.62 |
|  | 259 | 1.44 | 6.53 |
| 104 | 260 | 0.13 | 3.09 |
|  | 261 | 1.06 | 5.27 |
| 105 | 262 | 0.20 | 4.33 |
|  | 263 | 0.30 | 4.21 |
| 106 | 264 | −0.11 | 2.03 |
|  | 265 | 0.32 | 3.34 |
| 107 | 266 | −0.48 | 2.83 |
|  | 267 | −0.55 | 2.69 |
| 108 | 268 | −1.42 | 1.94 |
|  | 269 | −0.69 | 2.26 |
| 109 | 270 | −1.41 | 2.25 |
|  | 271 | −1.41 | 2.08 |
| 110 | 272 | −3.04 | 2.91 |
|  | 273 | −1.63 | 2.14 |
| 111 | 274 | −2.31 | 2.67 |
|  | 275 | −2.36 | 2.47 |
| 112 | 276 | −7.38 | 5.04 |
|  | 277 | −2.79 | 3.07 |
| 113 | 278 | −3.83 | 4.18 |

*Table 2 (continued)*

|     |     |        |       |
|-----|-----|--------|-------|
|     | 279 | −7.05  | 3.95  |
| 114 | 280 | −4.71  | 8.40  |
|     | 281 | −3.40  | 5.12  |
| 115 | 282 | −2.96  | 6.87  |
|     | 283 | 1.71   | 6.60  |
| 116 | 284 | −3.77  | 13.07 |
|     | 285 | −1.63  | 8.39  |
| 117 | 286 | −1.86  | 10.82 |
|     | 287 | −1.85  | 10.51 |
| 118 | 288 | −3.41  | 19.12 |
|     | 289 | −2.07  | 12.95 |
| 119 | 290 | −2.92  | 16.09 |
|     | 291 | −2.95  | 15.74 |
| 120 | 292 | −4.65  | 26.63 |
|     | 293 | −3.26  | 18.88 |
| 121 | 294 | −4.02  | 22.77 |
|     | 295 | −4.01  | 22.37 |
| 122 | 296 | −5.52  | 35.66 |
|     | 297 | −4.25  | 26.23 |
| 123 | 298 | −4.74  | 30.92 |
|     | 299 | −4.77  | 30.47 |
| 124 | 300 | −6.11  | 46.26 |
|     | 301 | −4.62  | 35.08 |
| 125 | 302 | −5.11  | 40.60 |
|     | 303 | −5.02  | 40.09 |
| 126 | 304 | −6.38  | 58.51 |
|     | 305 | −5.68  | 45.50 |
| 127 | 306 | −7.35  | 51.87 |
|     | 307 | −8.09  | 51.31 |
| 128 | 308 | −10.21 | 72.45 |
|     | 309 | −8.86  | 57.53 |
| 129 | 310 | −9.09  | 64.79 |
|     | 309 | 9.54   | 64.40 |
| 130 | 314 | −10.15 | 85.58 |
|     | 313 | −9.28  | 71.23 |

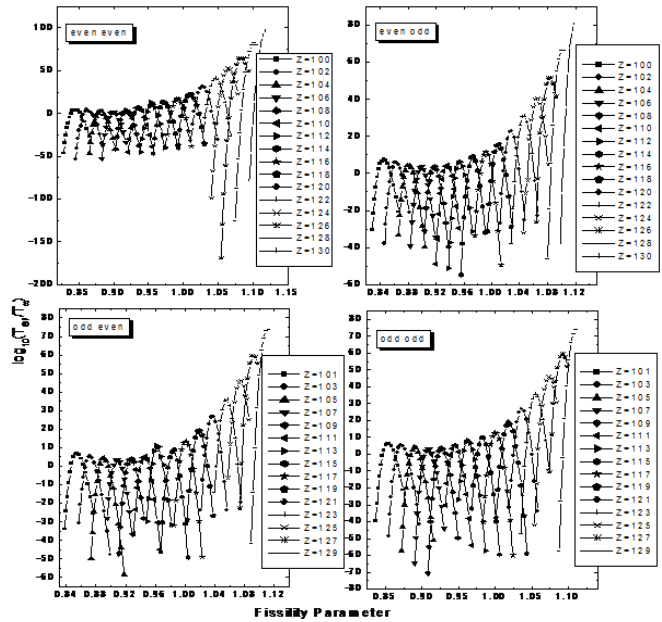

Fig. 2 – The ratio of spontaneous fission to alpha decay half-lives *versus* fissility parameter for even-even, even-odd, odd-even and odd-odd nuclei.

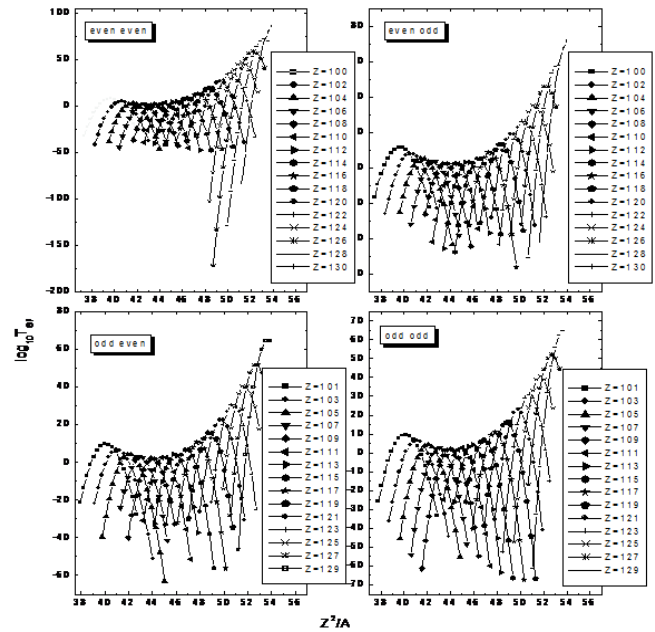

Fig. 3 – The spontaneous fission half-lives *versus* fissionability parameter ($Z^2/A$) for even-even, even-odd, odd-even, odd-odd isotopes.

We have also studied the plots connecting spontaneous fission half-lives against relative neutron excess, $I = (N-Z)/(N+Z)$ for even-even, even-odd, odd-even and odd-odd nuclei. Figure 4 represents the plot connecting the ratio of spontaneous fission to alpha decay half-lives against relative neutron excess for all types of nuclei. From these plots, it is seen that the curves converge at some points and then afterwards diverge. The isotopes in the first quadrant of this plot have alpha decay as the dominant mode of decay. So these isotopes will survive fission and can be synthesized and identified *via* alpha decay.

## 4. CONCLUSIONS

We have studied the alpha decay and spontaneous half-lives of the superheavy nuclei in the atomic range $Z = 100$–$130$ for even-even, even-odd, odd-even and odd-odd types. Any isotope of given superheavy nucleus which lies within the parabola of $\log T_{1/2}$ *versus* mass number plot will have higher half-life for spontaneous fission than alpha decay. Therefore by producing such isotopes of the superheavy nuclei, which have more half-life for spontaneous fission compared to alpha decay, one can detect the alpha particle and confirm the presence of such heavy nuclei. We presume that the present work on alpha decay favoured isotopes of superheavy nuclei may be a guide for future experiments.

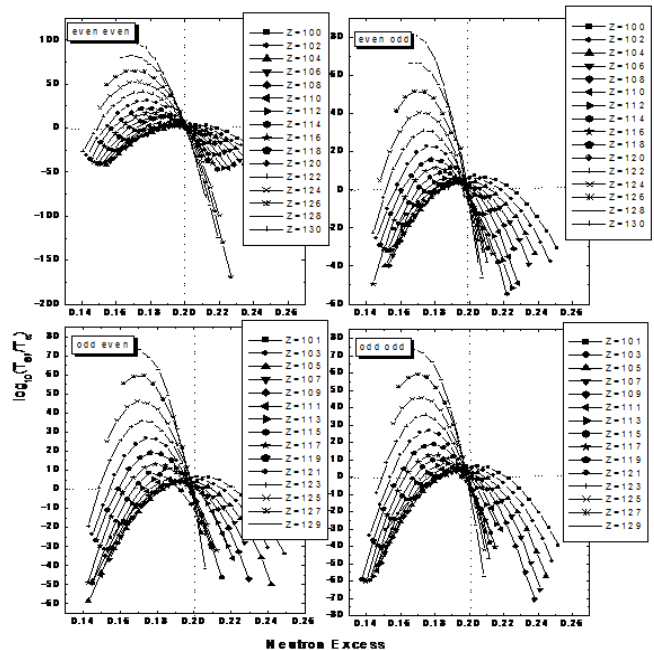

Fig. 4 – The ratio of spontaneous fission to alpha decay half-lives against relative neutron excess of even-even, even-odd, odd-even and odd-odd type nuclei.